\documentclass[prb,twocolumn,showpacs,superscriptaddress]{revtex4}
\usepackage{graphicx}
\usepackage{times}
\usepackage{bm}
\usepackage{hyperref}
\usepackage{amsmath, verbatim}
\newcommand{\ket}[1]{|#1\rangle}
\newcommand{\bra}[1]{\langle#1|}

\begin{document}

\title{Dimensional crossover in spin-orbit-coupled semiconductor nanowires with induced superconducting pairing}

\author{Tudor D. Stanescu}

\affiliation{Department of Physics, West Virginia University, Morgantown, WV 26506, USA}
\author{Roman M. Lutchyn}
\affiliation{Station Q, Microsoft Research, Santa Barbara, CA 93106-6105, USA}
\author{S. Das Sarma}
\affiliation{Condensed Matter Theory Center, Department of Physics, University of
Maryland, College Park, MD 20742, USA}

\begin{abstract}
We show that the topological Majorana modes in nanowires much longer than the superconducting coherence
length are adiabatically connected with discrete zero-energy states generically occurring in short nanowires. We
demonstrate that these zero-energy crossings can be tuned by an external magnetic field and are protected by
the particle-hole symmetry. We study the evolution of the low-energy spectrum and the splitting oscillations as
a function of magnetic field, wire length, and chemical potential, manifestly establishing that the low-energy
physics of short wires is related to that occurring in long wires. This physics, which represents a hallmark of
spinless $p$-wave superconductivity, can be observed in tunneling conductance measurements.
\end{abstract}

\pacs{03.65.Yz}
\maketitle

\section{Introduction}

The theoretical predictions~\cite{Fu2008, Zhang2008, Fu_PRB2009, Sato09, Sau2010, Alicea2010, Lutchyn2010, Oreg2010, Sau2010a, Potter2010, Lutchyn2011} that proximity effect induced by ordinary s-wave superconductors (SCs), along with spin-orbit coupling and Zeeman spin splitting, could give rise to topological superconductivity have led to an intensive experimental search for Majorana fermions. Following specific theoretical predictions~\cite{Lutchyn2010,Oreg2010}, a series of recent experimental papers~\cite{Mourik2012,Rokhinson2012,Deng2012, Das2012} have presented evidence for the existence of Majorana modes in quasi-1D semiconductor (SM) nanowires.  This excitement is further enhanced by the fact that these Majorana end modes can, in principle, be used to carry out fault--tolerant topological quantum computation~\cite{Nayak2008, AliceaBraiding}, as envisioned originally by Kitaev~\cite{Kitaev2001} more than 10 years ago.

Any observation of the Majorana mode in solid state materials is a rather important  experimental discovery, therefore it is legitimate to ask critically whether the recent experimental findings are truly consistent with the theoretical predictions for the elusive Majorana particle.
This is particularly important in view of the fact that the current experimental observations (except Ref. \cite{Rokhinson2012}) are based entirely on the existence of  zero--bias conductance peaks (ZBCPs) in the differential tunneling measurements, which represents a necessary condition~\cite{Sengupta2001,Law2009, Sau2010a, Flensberg, Fidkowski} for the existence of the Majorana mode. The sufficient condition necessitates an interference experiment establishing the non-Abelian nature of these modes, which has not yet been performed.
Since ZBCPs arise quite commonly in both SCs and SMs, it is of critical importance to carefully analyze the various experimental data to see whether the ZBCP is indeed consistent with the existence of the Majorana, or is arising from other, presumably more mundane, physical mechanisms~\cite{Franceschi, Rainis,Kells, Prada}.
In addition, in short wires with lengths comparable to the SC coherence length, it is commonly believed that the two end Majorana modes should hybridize and move away from zero bias~\cite{Cheng2009, Lin2012}.
The important practical question of fundamental significance addressed here concerns the issue of the shortest nanowire length consistent with the manifestation of a zero bias conductance peak indicating the presence of zero--energy modes in the underlying energy spectrum. This issue has become urgent because the original observation of the ZBCP in long ($> 2\mu$m ) InSb nanowires~\cite{Mourik2012} has recently been qualitatively reproduced in short ($< 0.5\mu$m) InAs nanowires~\cite{Das2012},
thus raising the important question of whether the ZBCPs in long and short wires are manifestations of the same qualitative physics or not.

The goal of the current work is to critically investigate the wire length dependence of the ZBCP in SC nanowires and to clearly identify the nature of the ZBCP in short wires and its possible relationship to the Majorana zero-energy modes emerging in long wires. We establish that, for appropriate values of the magnetic field, the lowest energy mode of the SC system is characterized by an adiabatic continuity as a function of wire length and that ZBCPs generated by this near-zero-energy mode may exist even for wire lengths comparable to the SC coherence length. Therefore, although  in short wires the whole notion of a topological phase with non-local  zero-energy Majorana modes becomes meaningless (i.e. anyons are strongly overlapping and cannot be manipulated independently), the mode characterized by zero-energy crossings associated with the ZBCPs corresponds to a pair of overlapping Majorana states and can be viewed as remnant Majorana physics carried over from the long-wire topological phase. We thus believe that the ZBCPs observed in Refs. \cite{Mourik2012} and \cite{Das2012} for long and short wires, respectively,  are adiabatically connected and, in some sense, are both manifestations of the predicted Majorana quasiparticles in topological quasi--1D superconductors~\cite{Lutchyn2010,Oreg2010}.
By solving numerically an effective tight-binding model for multiband SM nanowires with realistic parameters, we calculate the energy spectrum and the density of states as functions of the wire length and externally tunable parameters - magnetic field and chemical potential.
We show that robust zero--energy crossings associated with the remnant Majorana mode generically occur in short nanowires at discrete values of the  magnetic field ${\bm B}$.When isolated in the
parameter space, these zero-energy crossings are protected by
the particle-hole symmetry and represent a hallmark of spinless
$p$-wave superconductivity. 
With increasing wire length, the period of the zero--energy crossings and the amplitude of the energy splitting oscillations as function of Zeeman field or chemical potential decrease.
We demonstrate that these rather generic zero--energy crossings in short wires generate ZBCPs that may look similar to those produced by topological Majorana zero--energy modes due to the limited experimental resolution. 
The short wire low-energy remnant Majorana mode becomes the true topological Majorana bound state in the long wire limit.

\begin{figure}[tbp]
\begin{center}
\includegraphics[width=0.48\textwidth]{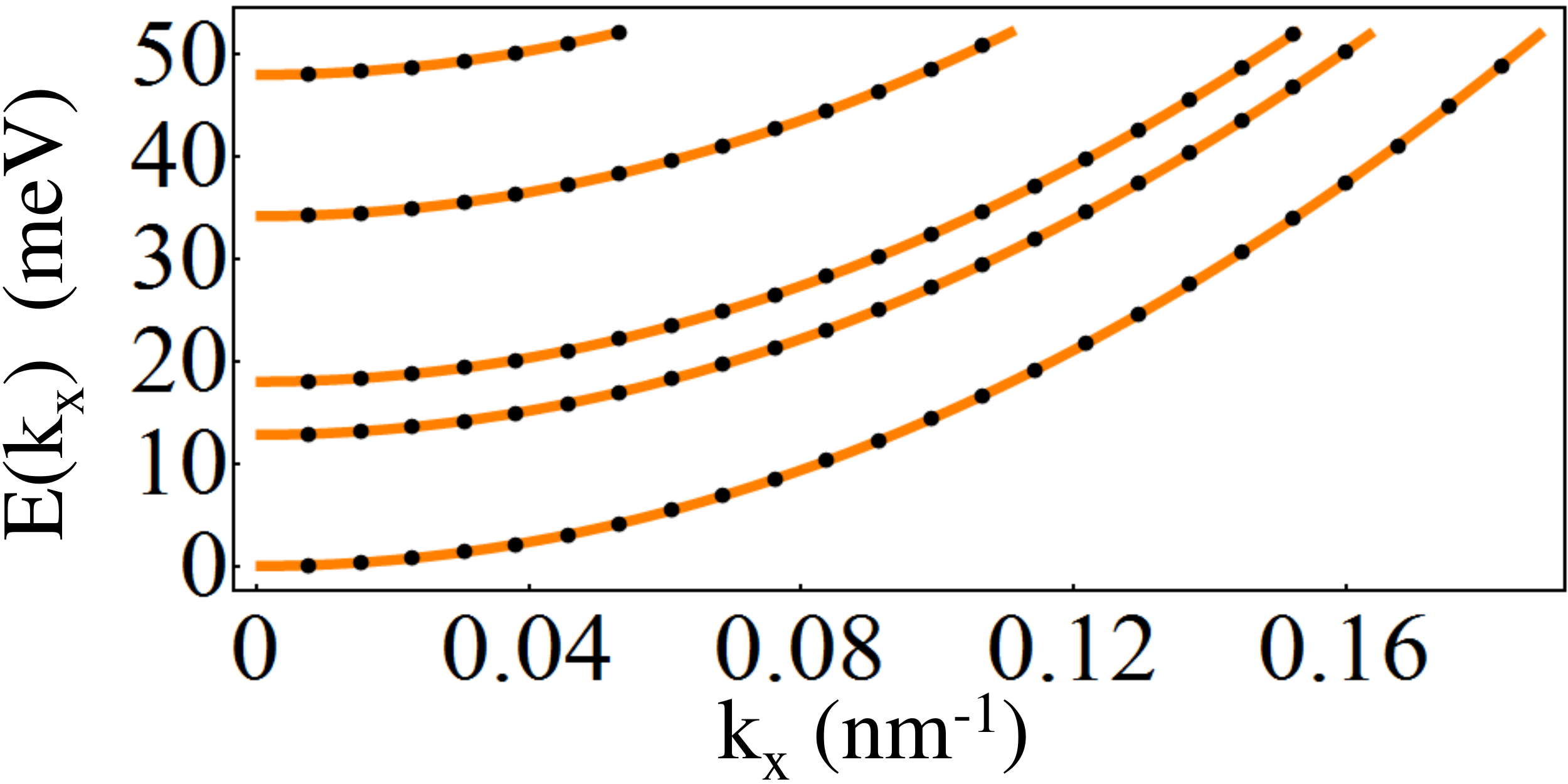}
\vspace{-4mm}
\end{center}
\caption{(Color online) Low-energy semiconductor spectrum in the non-superconducting phase (orange/light gray) corresponding to the effective parameters of Eq. (\ref{Hnm}). The black dots represent the discrete energy values for a short wire with $L_x\approx400$nm and periodic boundary conditions.}
\vspace{-2mm}
\label{Fig1}
\end{figure}

\section{Model}

 We consider a SM nanowire with rectangular cross section $L_y\times L_z = 50$nm$\times 60$nm and different wire length values $L_x$. In the limit of an infinite wire, $L_x\rightarrow\infty$, the Hamiltonian describing the nanowire reads \begin{align}
H^{\rm SM}_{\bm n \bm m}(k)=[\epsilon_{\bm n}(k) +\alpha_R k \sigma_y + \Gamma\sigma_x]\delta_{{\bm n \bm m}} - i \alpha q_{{\bm n \bm m}} \sigma_x, \label{Hnm_S}
\end{align}
where $k\!\equiv \!k_x$ is the wave number, $\sigma_i$ are Pauli matrices associated with the spin degree of freedom, and $\alpha_R=\alpha a$, is the strength of the Rashba spin-orbit coupling, with $a$ being the lattice constant. In Eq.~(\ref{Hnm}) ${\bm n} = (n_y,n_z)$ and  ${\bm m}=(m_y,m_z)$ label different confinement--induced sub--bands described by the transverse wave functions $\phi_n(y) \propto \sin( n_y\pi y/L_y)\sin( n_z\pi z/L_z)$, $\epsilon_{\bm n}(k)$ describes the SM spectrum without SO coupling, and $\Gamma= g^* \mu_B B/2$ is the external Zeeman field along the $x$-direction. The term containing $q_{{\bm n m}}$ represents the inter--band Rashba coupling\cite{Stanescu2011}  and is given explicitly in the Appendix.  The numerical values of the parameters correspond to InAs - effective mass $m_{\rm eff}= 0.026m_0$ and Rashba coefficient $\alpha_R=0.2$eV\AA. This defines spin-orbit length scale $l_{so}\equiv \hbar^2/m_{\rm eff}\alpha\approx 150$nm below which the spin-orbit coupling is effectively quenched. The low-energy spectrum of the wire is shown in Fig.~\ref{Fig1}. For a finite nanowire, the spectrum consists of discrete energy levels, as shown in Fig.~\ref{Fig1} for a short wire of length $L_x=400$nm.

\begin{figure}[tbp]
\begin{center}
\includegraphics[width=0.48\textwidth]{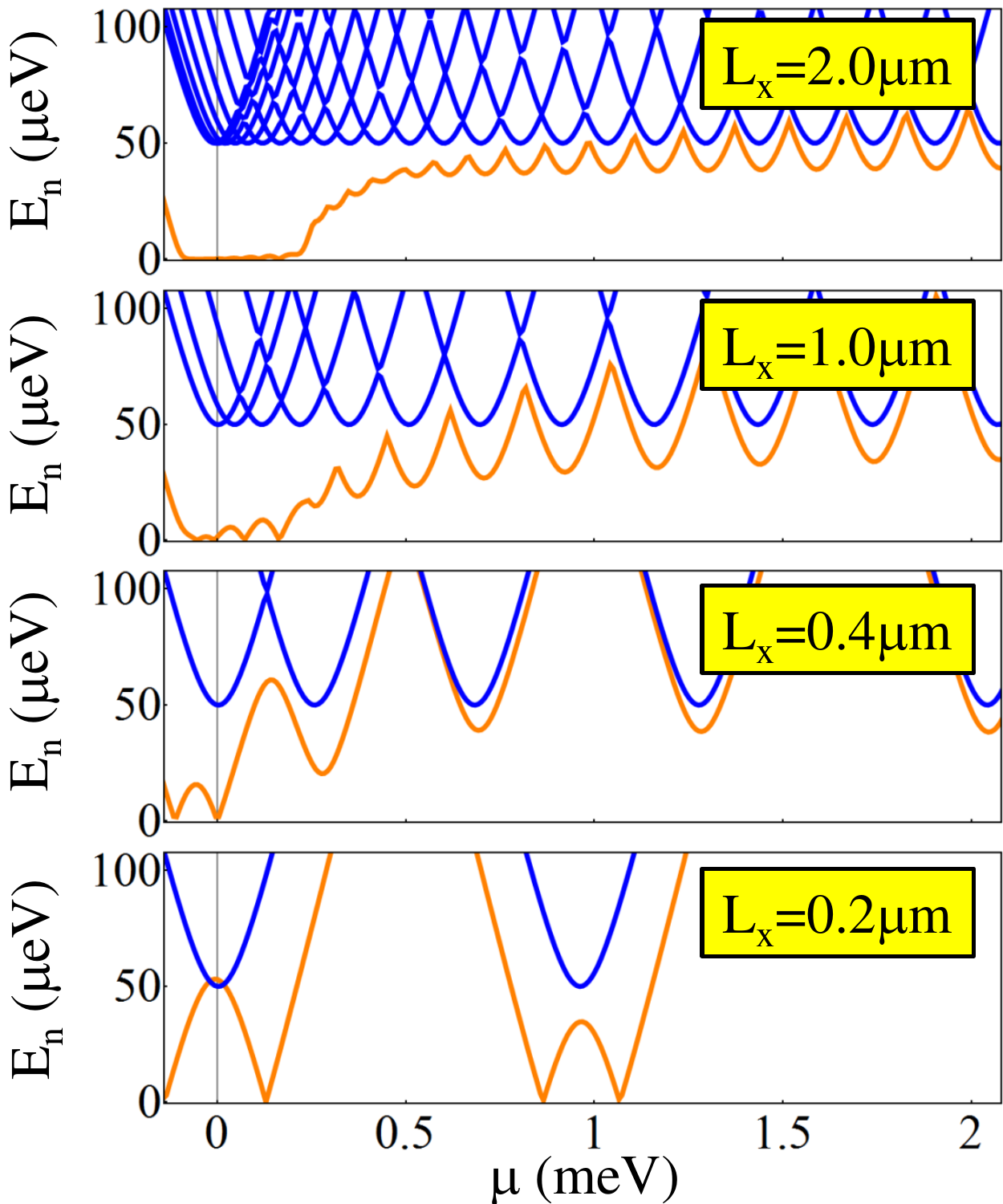}
\vspace{-4mm}
\end{center}
\caption{(Color online) Low energy BdG spectrum as function of the chemical potential for finite wires of different lengths. The blue (dark gray) lines correspond to a vanishing Zeeman field, while the orange (light gray) line represents the  lowest energy state for $\Gamma = 0.18$meV.  }
\vspace{-2mm}
\label{Fig2}
\end{figure}

Next, we consider the SM nanowire proximity-coupled to an s-wave superconductor. The superconductor can be described by the BCS density of states $\nu(E)=\nu_F \Theta(|E|-|\Delta_0|)\frac{E}{\sqrt{E^2-\Delta_0^2}}$ where $\nu_F$ and $\Delta_0$ are the normal density of states at the Fermi level and the SC energy gap, respectively. By integrating out the SC degrees of freedom and linearizing the frequency dependence, one arrives at an effective low-energy description of the system valid at energies $E \ll \Delta_0$~\cite{Stanescu2011} (see the Appendix). The corresponding  BdG Hamiltonian for quasi--1D nanowire reads
\begin{eqnarray}
 H_{{\bm n},{\bm m}}(k_x) &=& Z \left[\epsilon_{\bm n}(k_x) +\alpha_R k_x \sigma_y + \Gamma\sigma_x\right]\delta_{{\bm n \bm m}}\tau_z \nonumber \\
&+& i Z \alpha q_{{\bm n \bm m}} \sigma_x + \Delta_{\rm ind} \sigma_y \tau_y,      \label{Hnm}
\end{eqnarray}
where $\tau_i$ are  Pauli matrices associated with the particle-hole degree of freedom and we have used the basis $(u_{\uparrow}, u_{\downarrow}, v_{\uparrow}, v_{\downarrow})$ for the Nambu spinors.
In Eq. (\ref{Hnm}) the proximity--induced renormalization factor is $Z= (1+\gamma/\Delta_0)^{-1}$, where $\gamma=75\mu$eV is the effective SM--SC coupling, and the induced SC gap is $\Delta_{\rm ind} = \gamma\Delta_0/(\gamma+\Delta_0)= 50\mu$eV.
For these parameters, the level spacing between different $n_x$ states becomes larger than the SC gap $\Delta_0=150\mu$eV in wires with $L_x \leq 0.5\mu$m. In this regime, changing of the chemical potential leads to significant variations of the energy corresponding to the lowest BdG state and of the number of quasiparticle states within the SC gap $\Delta_0$. This behavior is illustrated in Fig. \ref{Fig2}. In the absence of a Zeeman field (blue/dark gray lines), the minima  of the BdG spectrum roughly correspond to the quantized energy levels $E_{n_x n_y n_z}=\mu$, with $(n_y, n_z)=(1,1)$, i.e., the lowest energy band in Fig. \ref{Fig1}, and different $n_x$ values. A similar behavior can be observed when the chemical potential is in the vicinity of other band minima, e.g., $\mu=18$meV $+ \Delta\mu$ for the band with $(n_y, n_z)=(2,1)$, plus extra contributions from the lower energy bands. In the presence of a Zeeman field, the energy of the lowest-energy state decreases and eventually vanishes at a certain $\mu$-dependent value of $\Gamma$. Note that, as a result of spin-orbit coupling,  states with low $n_x$ depend strongly on $\Gamma$, while high $n_x$ states are weakly $\Gamma$-dependent.

\begin{figure}[tbp]
\begin{center}
\includegraphics[width=0.48\textwidth]{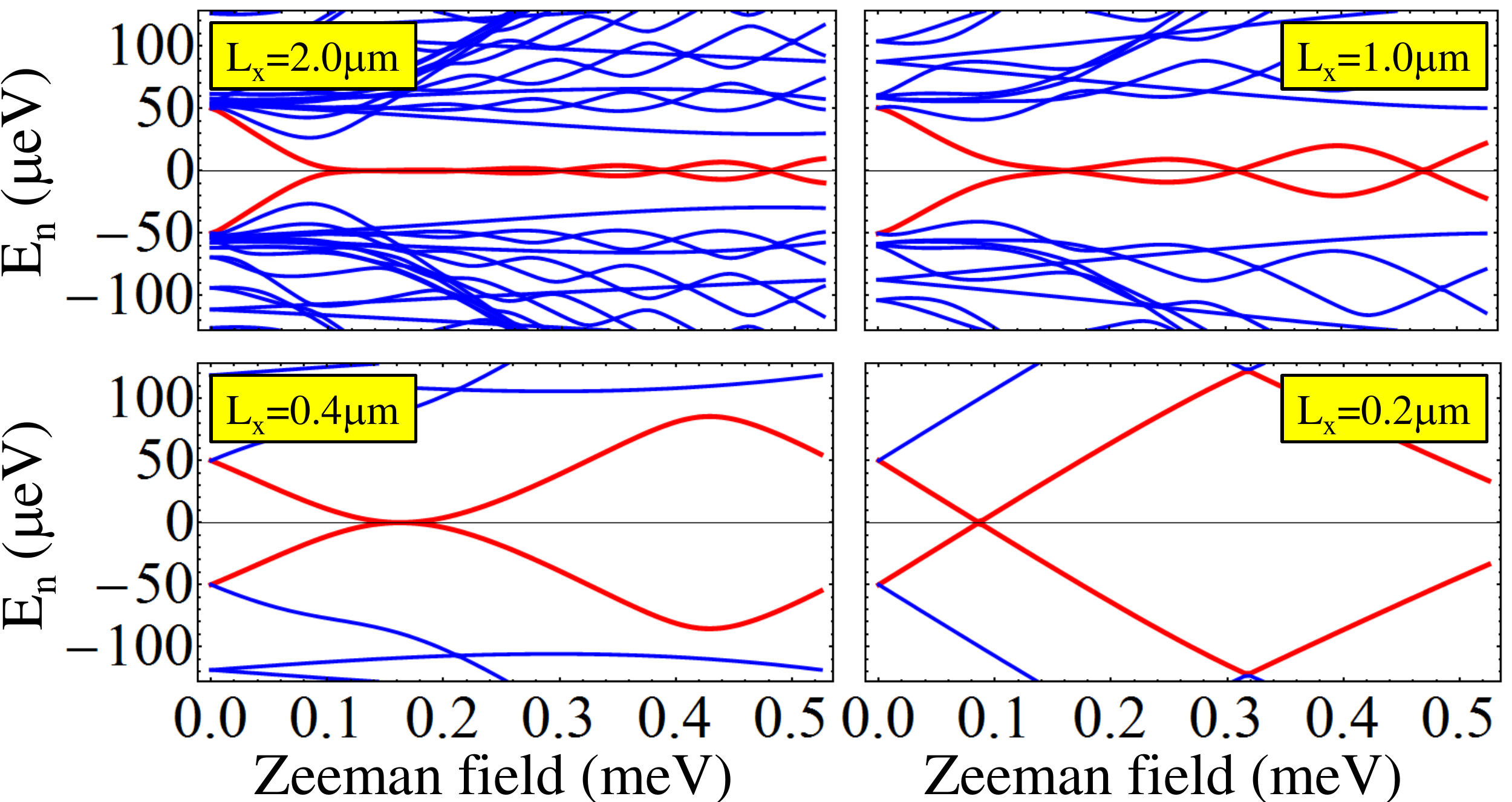}
\vspace{-4mm}
\end{center}
\caption{(Color online) Low energy BdG spectrum as function of the Zeeman field for finite wires of different lengths. In the limit $L_x\rightarrow\infty$ a Majorana zero mode appears above a critical field $\Gamma_c\approx 0.1$meV. In finite wires, the mode acquires a finite energy due to the overlap of the states localized at ends of the wire. In very short wires (e.g., $L_x=0.2\mu$m) the lowest energy state depends almost linearly on the Zeeman field. States characterized by different values of $n_x$ are coupled by the Rashba interaction and, consequently,  the dependence of their energy on $\Gamma$ is nonlinear. The chemical potential is $\mu=18$meV (bottom of the third band in Fig \ref{Fig1}).}
\vspace{-2mm}
\label{Fig3}
\end{figure}

\section{Numerical results and physical interpretation}

In the remainder of the paper, we focus on the experimentally--relevant parameter regime $L_x\sim \xi>l_{so}$ and contrast the properties of the system in this limit with the ones for a long nanowire ($L\gg\xi$).
The dependence of the quasiparticle spectrum on the applied magnetic field for several values of $L_x$ is shown in Fig.\ref{Fig3}. The lowest energy mode (red lines) is characterized by discrete zero--energy crossings that are robust against disorder, which we checked explicitly. In spinless superconductors, such isolated crossings are quite robust against perturbations due to the particle-hole symmetry. Indeed, consider $k\cdot p$ perturbation theory near a crossing point. The two zero-energy solutions $\Psi_0$ and $\Psi_1$ are related by particle--hole symmetry, $\Psi_1=\tau_x \Psi_0^*$. In order to open a gap at the crossing point, the off--diagonal matrix element has to be non--zero, $\bra{\Psi_0}V\ket{\Psi_1}\neq 0$, where $V$ is a generic perturbation that satisfies particle-hole symmetry $\tau_x V \tau_x=-V^T$.
However, using particle-hole symmetry we have $V_{01}=\bra{\Psi_0}V\ket{\Psi_1}=\int dx \Psi_0^* V \Psi_1=-\int dx \Psi_0^* \tau_x V^T \Psi_0^*=-\int dx \Psi_1 V^T \Psi_0^*=0$. Thus, particle-hole symmetry ensures the robustness of isolated zero-energy crossings. Another way of understanding the robustness of an isolated zero-energy crossing invokes fermion parity - one can show that the two zero-energy states $\Psi_0$ and $\Psi_1$ actually correspond to a different fermion parity~\cite{Fu_PRB2009}. However, the position of the zero-energy crossing point is non-universal and changes with the perturbation, since the diagonal matrix elements are non-zero $\bra{\Psi_0}V\ket{\Psi_0}=-\bra{\Psi_1}V\ket{\Psi_1}$. In order to get rid of the zero-energy crossings one has to bring another pair of zero-energy states to the same point in the parameter space.  Then, four states would hybridize with each other since two of them will now have the same fermion parity and eventually result in the avoided level crossings. This is illustrated in Fig. \ref{Fig3} for $L_x\approx 0.4 \mu$m. In this case, small variations of the chemical potential will result in either two close zero-energy crossings ($\Delta\mu <0$), or an avoided crossing ($\Delta\mu > 0$). However, these avoided level crossings would still be near-zero-energy states and may produce ZBCPs in experimental systems, which invariably have finite energy resolutions.

\begin{figure}
\begin{center}
\includegraphics[width=0.48\textwidth]{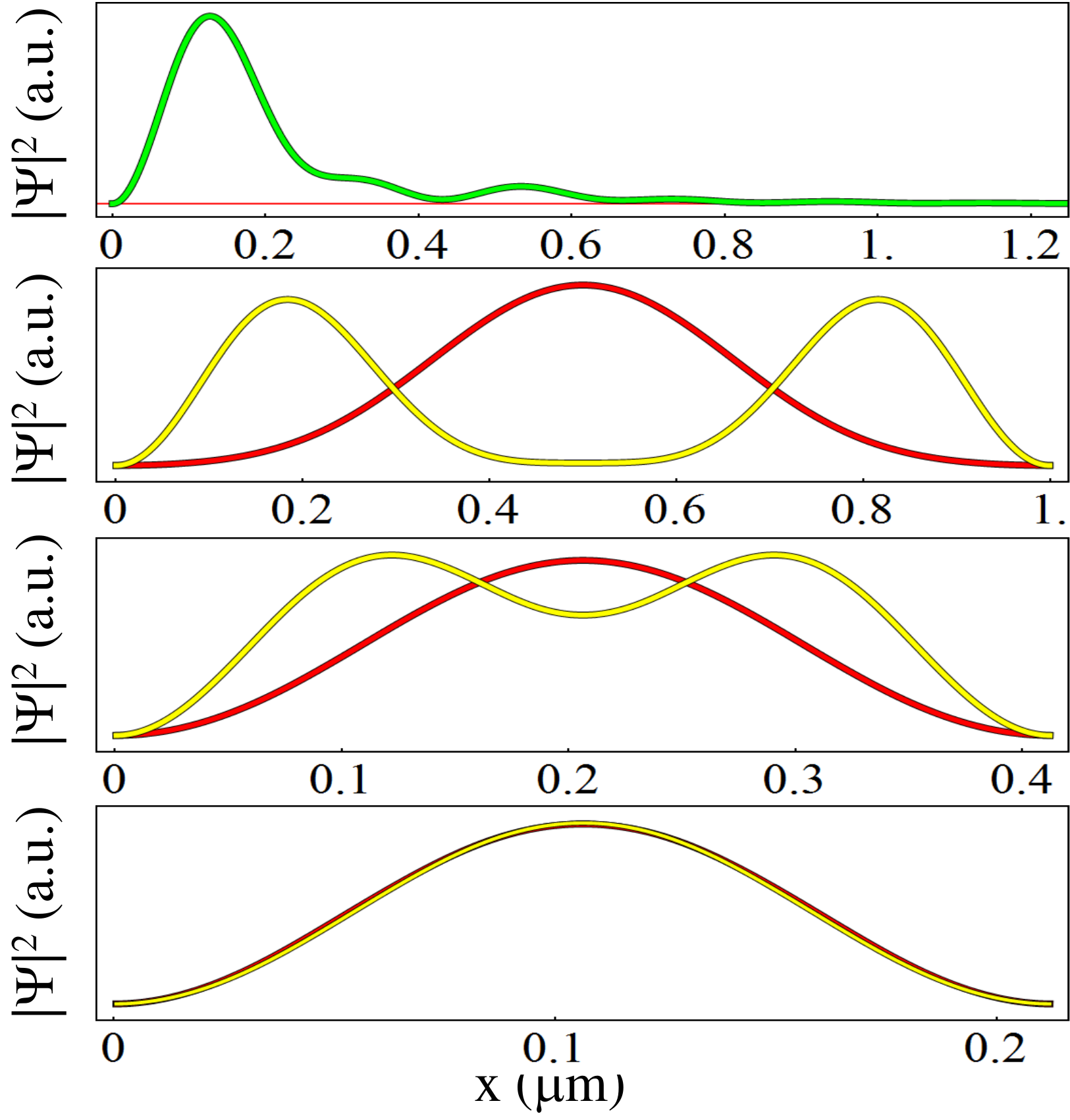}
\vspace{-4mm}
\end{center}
\caption{(Color online) Profiles of the lowest--energy states in nanowires of different lengths. Top panel: Majorana bound state localized near the end of a long wire ($L_x \gg \xi$). In the other panels the red (dark gray) lines correspond to $\Gamma=0.05$meV and the yellow (light gray) lines are for $\Gamma=0.18$meV (see Fig. \ref{Fig3}). Increasing the Zeeman field mixes states with different values of $n_x$ and generates modes that become localized near the end of the wire. This mechanism is absent in very short wires (bottom panel, ($L_x \ll \xi$)) due to the wide energy separation between the  quantized levels (see Fig \ref{Fig2}).}
\vspace{-2mm}
\label{Fig4}
\end{figure}

The emergence of zero--energy crossings in short wires $L_x\geq \xi>l_{so}$ is intriguing and one may ask the question whether it might be possible to use these zero--energy states for TQC. Indeed, one of the necessary ingredients for TQC is ground state degeneracy, which can be achieved hypothetically by fine--tuning. However, another important ingredient is the ability to manipulate the Majorana quasiparticles independently. Consider, for example, Kitaev’s lattice model\cite{Kitaev2001} at the special point when hopping $t$ is equal to gap $|\Delta|$. At this point, two zero-energy Majorana modes are  localized at the opposite ends of the chain. Thus, their manipulation, even in the limit of a short wire, would lead to non-Abelian braiding statistics. The situation at hand is different, however, because the quasi--Majorana modes are strongly overlapping, see Fig.~\ref{Fig4}, i.e. the anyons are strongly hybridized and their independent manipulation is not possible.

\begin{figure}
\begin{center}
\includegraphics[width=0.48\textwidth]{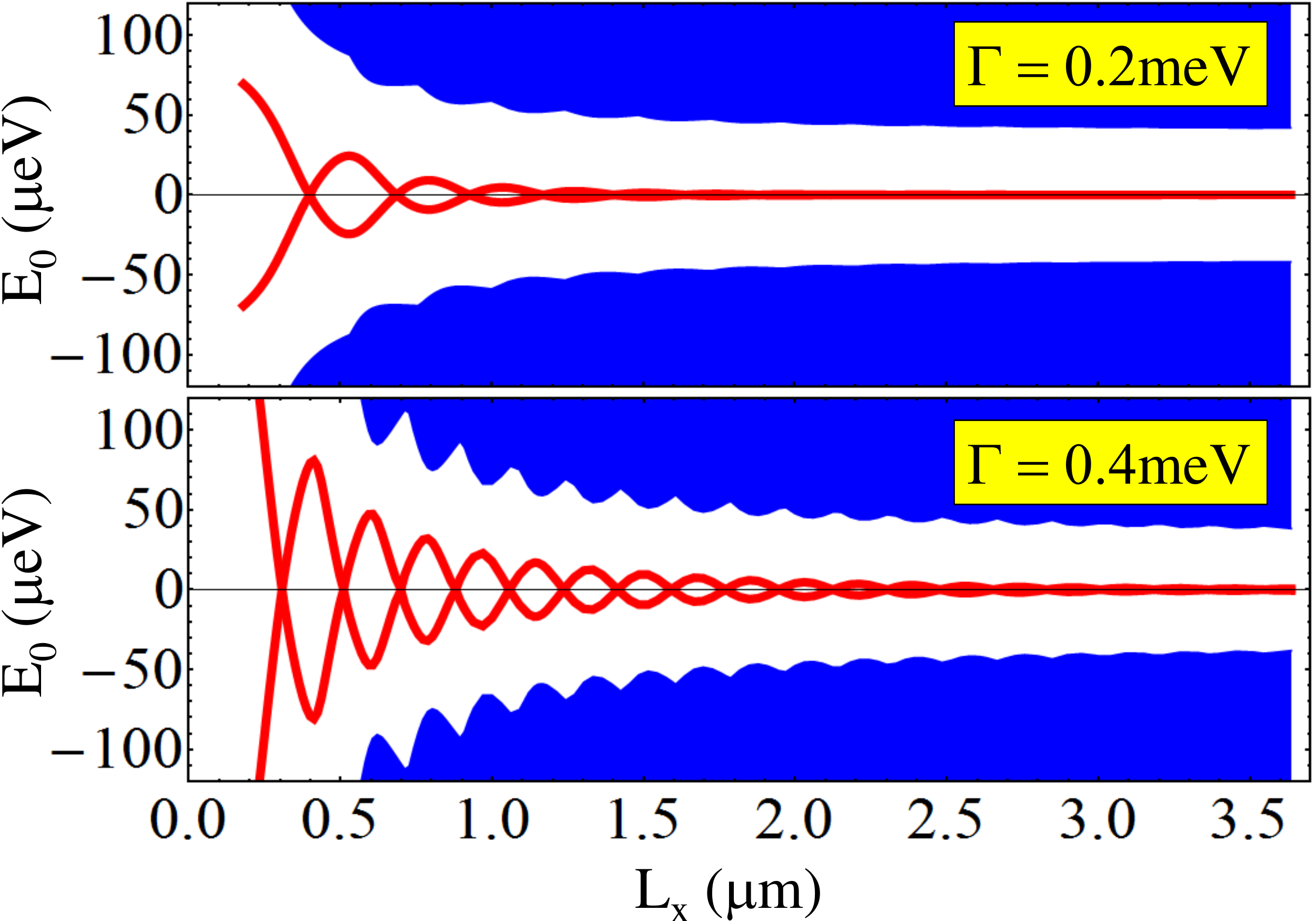}
\vspace{-4mm}
\end{center}
\caption{(Color online) Evolution of the lowest-energy mode with the size of the wire for different values of the Zeeman splitting $\Gamma=0.2$ meV and $\Gamma=0.4$ meV and chemical potential $\mu=0$. The shaded region corresponds to excited quasiparticle states. The size evolution of the low-energy spectrum corresponding to $\mu=13$meV is shown in the Appendix.}
\vspace{-2mm}
\label{Fig5}
\end{figure}

The evolution of the low-energy spectrum with the wire length clearly illustrates the adiabatic continuity of the near-zero-energy mode, as shown in Fig.~\ref{Fig5}. Consider first the long-wire limit $L_x\gg \xi$. Above the critical field $\Gamma_c >0.1$meV, the system is driven into a topological phase with Majorana zero-energy end states. In a finite system, the splitting energy $\delta E$ between Majorana modes has an oscillatory pre-factor, in addition to an exponentially-decaying envelope, $\delta E \propto \sin(k_F L_x)\exp(-L_x/\xi)$~\cite{Cheng2009}. Changing the system size or the magnetic field, which in turn changes $k_F$ and $\xi$, results in oscillations of the energy splitting, as shown in Fig.~\ref{Fig5} and Fig.~\ref{Fig3}.  With deceasing $L_x$, the number of oscillations within a given $\Gamma$ interval decreases, while their amplitude increases, so that the gap separating the lowest-energy mode from the excited states collapses, or $\delta E$ exceeds $\Delta_0$. At this point, $L_x\equiv L_c$, all remnant features of localized Majorana modes completely disappear. For the nanowires longer than $L_c$ there is adiabatic continuity of the spectrum indicating that there is no topological quantum phase transition between the $L_x\gg \xi$ and $L_x=L_c$ regimes. 

\begin{figure}[tbp]
\begin{center}
\includegraphics[width=0.48\textwidth]{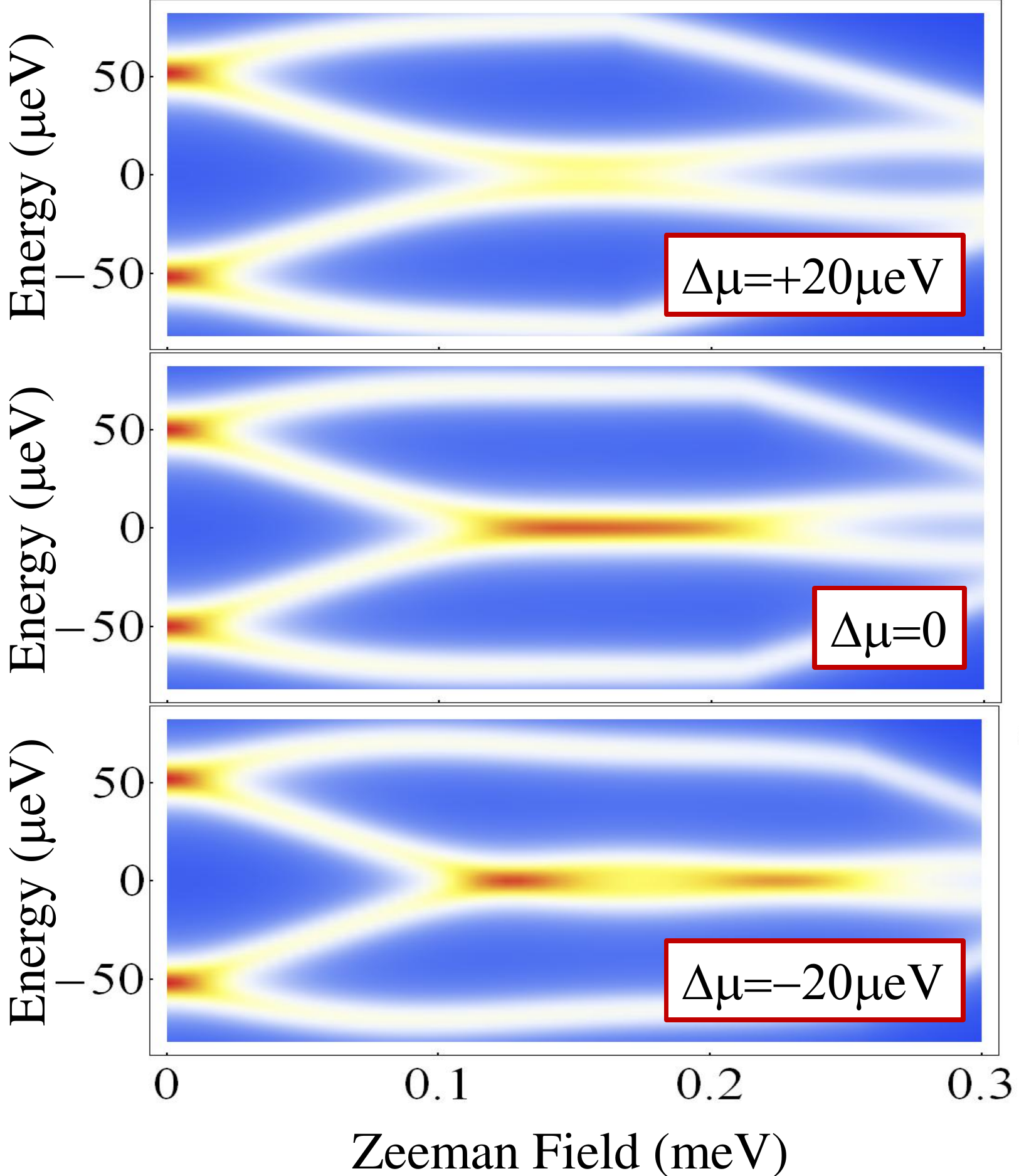}
\vspace{-4mm}
\end{center}
\caption{(Color online)  Nearly zero-energy peak in the density of states (DOS) of a short wire with $L_x=0.4\mu$m as function of the Zeeman field. The chemical potential is $\mu=18$meV. A decrease of the SC gap $\Delta_0$ with the increasing Zeeman field (by up to $60\%$) and a finite energy resolution of $10\mu$eV are included.}
\vspace{-2mm}
\label{Fig6}
\end{figure}

Our results presented in Figs.~\ref{Fig2}--\ref{Fig5} clearly establish that robust near-zero-energy modes are generic in quasi-1D nanowires in the presence of spin-orbit coupling, magnetic field, and SC pairing, in a wide range $L_x \geq \xi > l_{so}$.
One of the possible experimental implications of these findings is illustrated in Fig.~\ref{Fig6}. In the presence of a finite energy resolution, the near-zero-energy mode is converted into a continuous ZBCP as a function of the magnetic field, as observed experimentally. This behavior is reflected by the dependence of the calculated density of states (DOS) on the applied magnetic field (see Fig.~\ref{Fig6}). Note that the finite  width of the zero energy peak in the DOS is determined by temperature and coupling to the metallic leads. In short wires, where only a few states have energies inside the SC gap, this may be the dominant contribution. The apparent  ZBCP will eventually split off and may come back again at still higher fields.

\section{Conclusions}

We conclude by emphasizing that our findings have important implications for the current experiments probing the existence of Majorana modes in hybrid semiconductor structures.  Except for very short wires characterized by quantized level spacings much larger than $\Delta_0$ and $E_{\rm SO}$, the system supports an adiabatically continuous low--energy mode that smoothly crosses over, as the wire length increases, from a quasi--Majorana regime characterized by discrete zero--crossings and energy splitting oscillations to a zero--energy topological Majorana mode. This mode is generically associated with zero bias conductance peaks that, in a finite resolution measurement, extend over a finite magnetic field range.  This indicates that the recent observations in long InSb nanowires~\cite{Mourik2012} and in short InAs nanowires~\cite{Das2012} are adiabatically connected and are likely to be the expected signature for the predicted spinless p--wave superconductivity characterized by the existence of the Majorana quasiparticles in long nanowires.

This work is supported by Microsoft Q.

\appendix

\section{Low--energy effective model}

The realization of zero--energy Majorana bound states in solid state systems requires three key ingredients: i) strong spin--orbit coupling, ii) Zeeman splitting, and iii) superconductivity. In semiconductor wire--superconductor hybrid structures, these ingredients are provided by the spin--orbit interacting semiconductor, the external magnetic field, and the proximity induced superconductivity, respectively. The Hamiltonian that describes the quantum properties of the system has the generic form
\begin{equation}
H_{\rm tot} = H_{\rm SM} + H_{\rm Zeeman} + H_{\rm SC} + H_{\rm SM-SC},       \label{Htot}
\end{equation}
where different terms correspond to the semiconductor wire, the applied Zeeman field, the s-wave superconductor, and the semiconductor--superconductor coupling, respectively. The semiconductor term, which includes the spin--orbit coupling, is represented by the tight--binding Hamiltonian
\begin{eqnarray}
H_{\rm SM} &=& H_0 +H_{\rm SOI} = \sum_{{\bm i}, {\bm j}, \sigma} t_{{\bm i}{\bm j}}c_{{\bm i}\sigma}^{\dagger}c_{{\bm j}\sigma} -\mu \sum_{{\bm i}, \sigma} c_{{\bm i}\sigma}^{\dagger}c_{{\bm i}\sigma}  \nonumber \\
&+& \frac{i \alpha}{2}\sum_{{\bm i},{\bm \delta}}\left[ c_{{\bm i}+{\bm \delta}_x}^{\dagger}{\sigma}_y c_{{\bm i}} -  c_{{\bm i}+{\bm \delta}_y}^{\dagger}{\sigma}_x c_{{\bm i}} + {\rm h.c.} \right],  \label{Hsm}
\end{eqnarray}
where $H_0$, which includes the first two terms, describes nearest neighbor hopping on a simple cubic lattice with lattice constant $a$ with $t_{{\bm i}{\bm i}+{\bm \delta}} = -t_0$, where ${\bm \delta}$ are the nearest--neighbor position vectors. The hopping parameter can be expressed in terms of  the electron  effective mass in InAs as $t_0 = \hbar^2/(2m_{eff}a^2)$, with $m_{eff}=0.026m_0$, where $m_0$ is the bare electron mass.
 In Eq. (\ref{Hsm}) the last term represents the Rashba spin-orbit interaction (SOI), $c_{{\bm i}}^{\dagger}$ is a spinor $c_{{\bm i}}^{\dagger}=( c_{{\bm i}\uparrow}^{\dagger}, c_{{\bm i}\downarrow}^{\dagger})$ with $c_{{\bm i}\sigma}^{\dagger}$ being the electron creation operators with spin $\sigma$, $\mu$ is the chemical potential,   $\alpha$ is the Rashba coupling constant, and ${\bm \sigma}=(\sigma_x,\sigma_y,\sigma_z)$ are Pauli matrices. For a nanowire with rectangular cross section and dimensions $L_x \gg L_y\sim L_z$, the quantum problem corresponding to $H_0$ can be solved analytically and we obtain the eigenstates $\psi_{{\bm n} \sigma}({\bm i}) = \prod_{\lambda=1}^3 \phi_{n_\lambda}(i_\lambda) \chi_\sigma$, where ${\bm n}=(n_x, n_y, n_z)$ with $1\leq n_\lambda \leq N_\lambda$, $\chi_\sigma$ is an eigenstate of the ${\sigma}_z$ spin operator, and
\begin{equation}
\phi_{n_\lambda}(i_\lambda) = \sqrt{\frac{2}{N_\lambda+1}}\sin\frac{\pi n_\lambda i_\lambda}{N_\lambda+1},
\end{equation}
with $\lambda = x, y, z$ and $L_\lambda = a N_\lambda$, where $a$ is the lattice constant. Note that, for an infinite wire, the wave vector is a good quantum number, $n_x \rightarrow k_x$, and the corresponding eigenfunction becomes $\phi_{k_x}(x) =\sqrt{2/L_x} e^{i k_x x}$. The eigenvalues corresponding  to $\psi_{{\bm n} \sigma}$ are
 \begin{equation}
 \epsilon_{\bm n} \!=\! -2 t_0 \left( \cos\frac{\pi n_x}{N_x\!+\!1} \!+\!  \cos\frac{\pi n_y}{N_y\!+\!1} \!+\!  \cos\frac{\pi n_z}{N_z\!+\!1}-3\right)\!-\!\mu_0, \label{epsn}
\end{equation}
where ${\bm n}=(n_x,n_y, n_z)$ and the chemical potential  $\mu_0$ is measured from the bottom of the first band. For an infinite wire, the energy band corresponding to confinement--induced band ${\bm n}=(n_y, n_z)$ is given by
 \begin{equation}
 \epsilon_{\bm n}(k_x) \!=\! \frac{\hbar^2 k_x^2}{2 m_{eff}} -2 t_0 \left(\cos\frac{\pi n_y}{N_y\!+\!1} \!+\!  \cos\frac{\pi n_z}{N_z\!+\!1}-2\right)\!-\!\mu_0, \label{epsk}
\end{equation}

Since the number of degrees degrees of freedom in a finite wire is large (of the order $10^7$--$10^9$), yet Majorana physics is basically controlled by a reduced number of low--energy degrees of freedom  (of the order $10^3$--$10^4$), we project the problem into the low--energy subspace spanned by a certain number of low--energy eigenstates of $H_0$. We assume that only a few bands are occupied, so the low-energy subspace is defined by the eigenstates satisfying the condition $\epsilon_{\bf n} <\epsilon_{\rm max}$, where the cutoff energy  $\epsilon_{\rm max}$ is of the order $100$meV. Using this low-energy basis, the matrix elements of the SOI Hamiltonian can be written explicitly as \begin{eqnarray}
&~&\langle\psi_{{\bm n}\sigma}|H_{\rm SOI}|\psi_{{\bm n^\prime}\sigma^\prime}\rangle = \alpha \delta_{n_z n_z^\prime}\left\{ \frac{1-(-1)^{n_x+n_x^\prime}}{N_x+1}(i\hat{\sigma}_y)_{\sigma \sigma^\prime} \right. \nonumber \\
&~&~~~~~~~~~\times \left. \frac{\sin\frac{\pi n_x}{N_x+1}\sin\frac{\pi n_x^\prime}{N_x+1}}{\cos\frac{\pi n_x}{N_x+1}-\cos\frac{\pi n_x^\prime}{N_x+1}}\delta_{n_y n_y^\prime} - [x \Leftrightarrow y] \right\}, \label{HSOI}
\end{eqnarray}
where the second term in the parentheses is obtained from the first term by exchanging the  $x$ and $y$ indices. Note that the SOI Hamiltonian has the structure $H_{\rm SOI} = H_{\rm SOI}^x +H_{\rm SOI}^y$, where the first term represents the intra-band Rashba coupling, while $H_{\rm SOI}^y$ couples bands with different $n_y$ indices. For an infinite wire, the first term in (\ref{HSOI}), representing the intra--band contribution, becomes
\begin{equation}
\langle H_{\rm SOI}^x\rangle_{{\bm n n}^\prime} = \alpha_R k_x \delta_{{\bm n n}^\prime}\sigma_y,  \label{HSOIx}
\end{equation}
where $\alpha_R=\alpha a$. In the numerical calculations we use  $\alpha_R=0.2$eV\AA).
The inter--band spin--orbit coupling corresponding to the second term in (\ref{HSOI}) has the form
\begin{equation}
\langle H_{\rm SOI}^y\rangle_{{\bm n n}^\prime} = -i \alpha q_{{\bm n n}^\prime}\sigma_x,  \label{HSOIy}
\end{equation}
with
\begin{equation}
q_{{\bm n n}^\prime} = \frac{1-(-1)^{n_y+n_y^\prime}}{N_y+1}\frac{\sin\frac{\pi n_y}{N_y+1}\sin\frac{\pi n_y^\prime}{N_y+1}}{\cos\frac{\pi n_y}{N_y+1}-\cos\frac{\pi n_y^\prime}{N_y+1}}\delta_{n_z n_z^\prime}.  \label{qnn}
\end{equation}

The second ingredient for realizing Majorana fermions in semiconductor nanowires is represented by the Zeeman field. We consider that the Zeeman splitting $\Gamma$ is generated by applying a magnetic field oriented along the wire (i.e., along the $x$-axis),  $\Gamma = g^* \mu_B B_x /2$. The corresponding matrix element in the low-energy basis are
\begin{equation}
\langle\psi_{{\bm n}\sigma}|H_{\rm Zeeman}|\psi_{{\bm n^\prime}\sigma^\prime}\rangle =\Gamma \delta_{{\bm n}{\bm n}^\prime} \delta_{\bar{\sigma} \sigma^\prime}, \label{HZeemannn}
\end{equation}
where $\bar{\sigma}=-\sigma$. Adding together these contribution, the effective Hamiltonian describing the low--energy physics of the semiconductor nanowire in the presence of a Zeeman field becomes
\begin{equation}
H^{\rm SM}_{{\bm n n}^\prime}(k_x)=[\epsilon_{\bm n}(k_x) +\alpha_R k_x \sigma_y + \Gamma\sigma_x]\delta_{{\bm n n}^\prime} - i \alpha q_{{\bm n n}^\prime} \sigma_x,
\end{equation}
where $\epsilon_{\bm n}(k_x)$ is given by Eq. (\ref{epsk}) and $q_{{\bm n n}^\prime}$ by Eq.  (\ref{qnn}). For a finite wire, the bare energy is given by the expression in Eq.  (\ref{epsn}), while the spin--orbit contribution corresponds to Eq. (\ref{HSOI}).

\begin{figure}
\begin{center}
\includegraphics[width=0.48\textwidth]{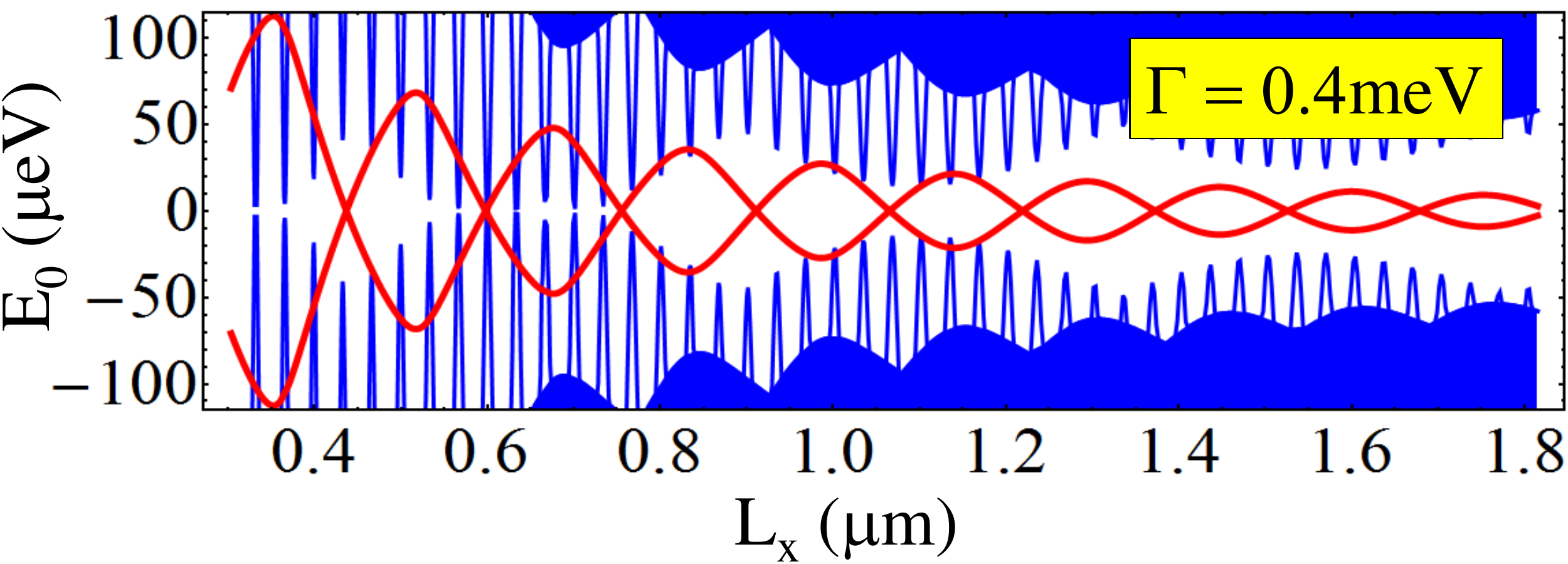}
\vspace{-4mm}
\end{center}
\caption{(Color online) Evolution of the quasi--Majorana mode with the size of the wire for a Zeeman splitting $\Gamma=0.4$ meV and chemical potential $\mu=13$meV corresponding to two partially occupied bands. The rapidly oscillating blue line represents regular Andreev bound states associated with the low--energy occupied band. In short wires, the energy of these Andreev bound states may vanish as a function of the Zeeman field, but they are not adiabatically connected to the topological Majorana bound states that emerge in long wires.}
\vspace{-2mm}
\label{Fig1S}
\end{figure}

The third key ingredient is the proximity-induced superconductivity (SC). As a result of the proximity to the s-wave superconductor, a pair potential $\Delta$ is induced in the nanowire and the energy scale for the quantum states in the semiconductor are renormalized. To account for this  effect, we integrate out the SC degrees of freedom and incorporate them as a surface self--energy term of the form \cite{SLDS}
\begin{align}\label{eq:Sigma_clean}
\Sigma(\omega)=-\gamma\left[ \frac{\omega + \Delta_0 \sigma_y\tau_y}{\sqrt{ \Delta_0^2-\omega^2}}+\zeta \tau_z\right],
\end{align}
where $\gamma=0.3$meV is the effective SM-SC coupling, $\tau_x$ and $\tau_z$ are Pauli matrices in the Nambu space, $\Delta_0=1.5$meV is the pair potential of the bulk SC, and $\zeta$ is a proximity-induced shift of the chemical potential. In the present calculations we take $\zeta=0$. Within the static approximation $\sqrt{ \Delta_0^2-\omega^2} \rightarrow \Delta_0$, the self-energy becomes $\Sigma(\omega)\approx -\gamma\omega/\Delta_0  - \gamma \sigma_y\tau_y$
and the low-energy physics of the SM nanowire with proximity-induced SC can be described by an effective Bogoliubov-de Gennes Hamiltonian.
This approximation is valid, strictly speaking, at energies much lower than $\Delta_0$, but represents a very good approximation even for $E\sim\Delta_0/2$. Explicitly, the matrix elements of the effective BdG Hamiltonian can be written as
\begin{eqnarray}
 H_{\rm BdG}({\bm n},{\bm n}^\prime) &=& Z \left[\epsilon_{\bm n} \delta_{{\bm n n}^\prime} + \Gamma \sigma_x \delta_{{\bm n}{\bm n}^\prime} + \langle H_{\rm SOI}^x\rangle_{{\bm n n}^\prime} \right]\tau_z \nonumber \\
&+& Z \langle H_{\rm SOI}^y\rangle_{{\bm n n}^\prime} + \Delta \sigma_y \tau_y,  \label{Hbdg}
\end{eqnarray}
where ${\bm n} =(n_x, n_y, n_z)$ are quantum numbers for the nanowire states in the absence of spin--orbit coupling,
$\epsilon_{\bm n}$ are the corresponding energies, $\Gamma$ is the Zeeman splitting, and $ \langle H_{\rm SOI}^x\rangle_{{\bm n n}^\prime}$ and $ \langle H_{\rm SOI}^y\rangle_{{\bm n n}^\prime}$  are matrix elements for the intra--band and inter--band  Rashba spin--orbit coupling, respectively. Note that energy scale for the SM nanowire is renormalized by a factor $Z= (1+\gamma/\Delta_0)^{-1}$ due to the SC proximity effect. This renormalization is determined by the term in the self-energy (\ref{eq:Sigma_clean}) that is proportional to $\omega$ (in the static approximation). The pairing term in Eq. (\ref{Hbdg}) is derived from the corresponding contribution to the self-energy (\ref{eq:Sigma_clean}) and is proportional to the induced pair potential $\Delta=\gamma\Delta_0/(\gamma+\Delta_0)=250\mu$eV. For an infinite wire, Eq. (\ref{Hbdg}) becomes
\begin{eqnarray}
 H_{{\bm n},{\bm n}^\prime}(k_x) &=& Z \left[\epsilon_{\bm n}(k_x) +\alpha_R k_x \sigma_y + \Gamma\sigma_x\right]\delta_{{\bm n n}^\prime}\tau_z \nonumber \\
&+& i Z \alpha q_{{\bm n n}^\prime} \sigma_x + \Delta \sigma_y \tau_y,  \label{Hnn}
\end{eqnarray}
where ${\bm n} =(n_y, n_z)$ labels the confinement--induced bands.
The effective BdG Hamiltonian described by Eq. (\ref{Hbdg}) (for a finite system) or Eq. (\ref{Hnn}) (for an infinite wire) is diagonalized numerically.

{\em Quasi--Majorana versus regular Andreev bound states}. We emphasize that the adiabatic connection between the quasi--Majorana mode in short wires and the topological Majorana mode is, in general, nontrivial. As illustrated in Fig. \ref{Fig1S}, in short wires, in addition to the quasi--Majorana mode, there are other low--energy Andreev bound states that may have vanishing energy at specific values of $L_x$ and Zeeman splitting. However, in long wires these modes will be characterized by a finite energy gap, while the energy of the Majorana mode will vanish. We note that these regular Andreev bound states are associated with the lower--energy occupied bands, in contrast with the Majorana (or quasi--Majorana) mode, which is always associated with the top occupied band.  Experimentally, the contributions arising from these types of low--energy states could be disentangled by varying the effective length of the wire (e.g., using a gate potential): the energy of quasi--Majorana mode will show weak dependence on the effective wire length, in sharp contrast with these regular Andreev bound states associated with the low--energy bands, which have energies that depend dramatically on the length of the wire.  

\bibliography{ShortNWref}

\end{document}